# Computer Simulation of 3-D Finite-Volume Liquid Transport in Fibrous Materials: a Physical Model for Ink Seepage into Paper

Reza Farrahi Moghaddam and Fereydoun Farrahi Moghaddam and Mohamed Cheriet

*Synchromedia Laboratory for Multimedia Communication in Telepresence,*
*École de technologie supérieure, Montreal, (QC), Canada H3C 1K3*
*Tel.: +1-514-396-8972*
*Fax: +1-514-396-8595*
*imriss@ieee.org, rfarrahi@synchromedia.ca, ffarrahi@synchromedia.ca,*
*mohamed.cheriet@etsmtl.ca*

**Abstract**

A physical model for the simulation ink/paper interaction at the mesoscopic scale is developed. It is based on the modified Ising model, and is generalized to consider the restriction of the finite-volume of ink and also its dynamic seepage. This allows the model to obtain the ink distribution within the paper volume. At the mesoscopic scale, the paper is modeled using a discretized fiber structure. The ink distribution is obtained by solving its equivalent optimization problem, which is solved using a modified genetic algorithm, along with a new boundary condition and the quasi-linear technique. The model is able to simulate the finite-volume distribution of ink.

## 1 Introduction

Historical documents and manuscripts are snapshots of the cultural past and the history of civilizations [2, 5]. They cannot be replaced by their modern counterparts, because marks or strokes on them that seem meaningless may reveal an important relationship to thought that would be unreachable in a transliteration. The task of preserving these cultural pearls is of great importance, therefore, as they are valuable resources that have survived threats to their existence many times over throughout the ages. However, they have become degraded by environmental factors, such as humidity and temperature, and by the aging process. Usually, the degree of the physical degradation is so great that other sources of degradation, such as that potentially caused by the imaging systems themselves, can be virtually ignored, especially since the camera-based imaging systems in use today in cultural archiving centers are of such high quality. Therefore, the main problem with old documents is physical degradation.

At the same time, understanding historical manuscripts and extracting text and textual relationships from them is a challenging task [34, 36, 17]. Because of the degradation phenomenon, the ink may be faded or some parts of the text may be missed. Also, the background intensities could be very complex, and interfere with the analysis process. If a good understanding of the physical phenomena that control the interaction of paper and ink is not available, the tasks of preservation and understanding cannot be accomplished.

A common phenomenon encountered in old documents is bleed-through [38, 13, 42, 22, 15, 14, 9, 37, 20, 18, 21, 19], which results in the appearance of interfering patterns and signatures on one side of the document originating from the ink strokes on the other side of it. The main problem with bleed-through signatures is that they resemble text-like patterns because their origin is the text patterns on the other side of the paper. Not only does this reduce our ability to understand and process the document images, it also makes it difficult to enhance and restore these images. Therefore, features such as orientation, color, and regional and local features should be used to identify these patterns, which may also vary from one document image to another. It soon becomes clear that modeling this phenomenon, and, more accurately, modeling the ink-paper interaction, plays a central role in the processing of degraded document images.

Understanding the ink seepage phenomenon is a challenging task, but it is an important requirement in document-related processing. This phenomenon can be considered as a fluid flow problem through a porous medium. The interaction of fluid with porous media is not limited to paper and ink, of course. From water transport in the soil to the oil industry [35, 39, 11, 41, 6], this interaction governs many critical and important physical problems. The application that is the most similar to the ink-paper problem is fabric printing. There is a big difference between these two applications, however. In fabric printing, the amount of ink involved is almost unlimited, and instead, gravity, time, and other parameters control the extent of ink seepage into the fabric structure. In contrast, in the ink-paper case, the amount of ink is the main constraint that determines the extent of ink propagation into the paper. Due to the limited amount of ink/water and other restrictions in the ink-paper problem, many of the approaches and methods that are based on stationary flow of the fluid, or reservoir-based transition flow, are not applicable.

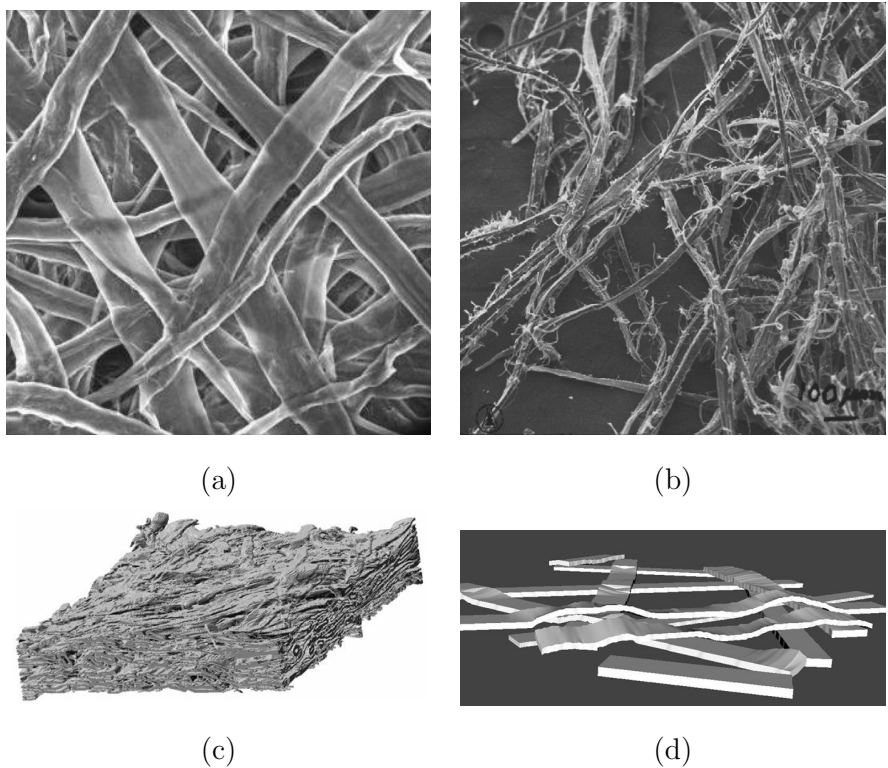

Fig. 1. a) and b) X-ray images of paper structures (courtesy of [4]). c) X-ray images of a piece of wood (courtesy of [4]). d) A fiber structure, which resembles a real structure, generated as a domain for numerical simulation (courtesy of [40]).

There are two main approaches to addressing the seepage phenomenon: physical experiments and numerical simulations. Although physical experiments seem to be more accurate, they are quite expensive, and the extraction of parameters of interest from the bulk of the physical material is difficult, in some cases introducing a large error. In contrast, thanks to remarkable developments in computational fluid dynamics, and also in computational hardware, numerical simulations now very nearly approach physical experiments in terms of accuracy. At the same time, the extraction of information from the output of numerical calculations is easy and straightforward.

The porous nature of paper is the main cause of ink seepage, as paper is usually made up of a collection of cellulose fibers. Figures 1(a) and 1(b) show two close-up X-ray images of paper fiber structures. The image on left is an example of newsprint paper. For the sake of comparison, the fiber structure of wood is also provided, in Figure 1(c). As can be seen from the figure, a number of imperfections are introduced into the fiber structure in the process of paper-making. Many works have used numerically generated fiber structures to simulate the fluid and solid interaction [40]. A sample structure is shown in Figure 1(d).

The seepage phenomenon can be analyzed at different scales: macroscopic, mesoscopic, microscopic, and nanoscopic. The macroscopic scale is the most practical of these; however, it requires that the physics of the problem be known, in terms of some nonlinear diffusion coefficient. The microscopic scale can be used to analyze a single fiber, but the computational complexity of processing a fiber structure would be tremendously high. The nanoscopic scale enables a detailed study of the interactions within a single fiber. But, it is at the mesoscopic scale [16] that individual fibers and fiber structure can be observed at the same time. Analysis at this scale could be undertaken independently of the other scales, because, in addition to the possibility of observing the majority of the interactions, it offers the option of developing a model using many different approaches. An example of fluid saturation in a fiber structure, obtained in [4], is presented in Figure 2. A brief description of some of these approaches is provided in section A.

In this work, a realistic model at the mesoscopic scale is developed. The new model is able to solve finite-volume fluid problems, which are critically important in the study of ink-paper interaction. To the best of our knowledge, no such model exists in the literature. The proposed model considers the individual fibers in the paper structure and follows the ink movement either within or between these fibers. In order to build this model, a new energy functional and a new boundary condition are introduced to model the finite-volume constraint. Also, a modified numerical optimization technique, based on the genetic algorithm (GA), is introduced to reduce computational cost. In this technique a quasi-linear approach is used to convert the nonlinear nature of the problem into a quasi-linear one which reduces that cost drastically.

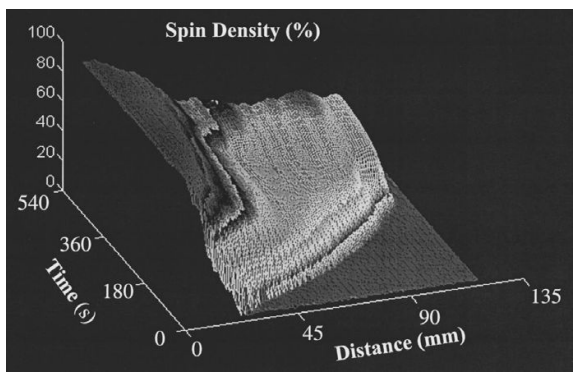

Fig. 2. Saturation profiles of the liquid in a fiber assembly (from [4]).

The organization of the paper is as follows. In section A, a review of the various models used to study the propagation of fluid in the porous medium is presented. The problem statement is given in section 4. The process of building the fiber structure at the mesoscopic scale is discussed in section 5. The formulation of the model, and especially the development of finite-volume behavior, is presented in section 6. Then, in section 7, two methods to solve the proposed model are provided, based on a global optimizer and a modified GA respectively. (We study various situations in this section.) Finally, our conclusion and some prospects for future research are presented in section 8.

## 2 Notation

Here, following [24], a few of the notations used in the paper are introduced:

- $x$ and $y$: the coordinates that describe the paper surface. For 3D models, $z$ represents the depth coordinate.
- $\phi$: the fiber field, which represents the fiber structure. If a point belongs to a fiber, $\phi$ is 1 at that point, otherwise $\phi$ is 0.
- $\sigma$: the ink field, which is 1 at a point if ink exists at that point, and is 0 otherwise. Another variable related to the ink is $\hat{\sigma}$. $\hat{\sigma}$ is the same as $\sigma$, except that its values are 1 and $-1$ respectively.

## 3 The modified Ising Model

The modified Ising models have been used in the study of solid-fluid interaction at mesoscopic scale [24, 27, 26, 31, 32, 25]. It assumes a lattice-based porous representation for the solid structure. Instead of modeling the interaction using complex transport coefficients of the macroscopic equations, it uses simple energy exchange mechanisms among the neighbors in the lattice. At this scale, the paper is almost equivalent to a fiber structure which consisting of a collection of fibers. Usually, the fiber structure is discretized into a collection of small cells. The material properties are assigned to the cells and to the mechanisms that control the interaction between neighboring cells. For example, a 3D model has been proposed in [27] to study the propagation of ink in the fabric to study the wetting process along a fiber bundle. The model is based on an energy functional:

$$E_{t,0} = \sum_i \left( E_{g,i} + E_{c,i} + E_{a,i} \right) \tag{1}$$

where $E_{t,0}$ is the total interaction energy of the ink/paper material, in which $E_{g,i}$, $E_{c,i}$, $E_{a,i}$ are the gravitational, cohesive, and adhesive energy terms respectively:

$$E_{g,i} = G_g \hat{\sigma}_i z_i, \quad E_{c,i} = -c_1 \hat{\sigma}_i S_{1,i}, \quad E_{c,i} = -A_1 \hat{\sigma}_i F_{1,i}$$

where $G_g$ is the gravity constant, $\hat{\sigma}_i$ is the ink state of the cell $i$ (also called the spin), and $z_i$ is the vertical position of that cell. $\hat{\sigma}_i$ is 1 if the cell is filled with the fluid, and -1 if the cell in empty. $S_{1,i}$ is the ink state of the neighbors:

$$S_{1,i} = \sum_{j \in N_{1,i}} \hat{\sigma}_j,$$

where $N_{1,i}$ is the set of cells that are the immediate neighbors of cell $i$: $N_{1,i} = \{j | i \text{ and } j \text{ share a wall}\}$. A schematic diagram of $N_1$ is shown in Figure 3(a). The term $\sum_i E_{c,i}$ stands for the cohesive energy between fluid particles. The last term in the total energy (1) is the adhesive term, and accounts for the adhesive interaction between the fluid and solid particles, in which

$$F_{1,i} = \sum_{j \in N_{1,i}} \phi_j,$$

where $\phi_j$ is the state of cell $i$ with respect to the solid material. If the cell is a part of a fiber, then $\phi_j$ is 1, and if not, it is -1. As the fiber structure is stationary, the field $\phi$ is also stationary and fixed. The solution is an ink distribution $\hat{\sigma}$ that minimizes $E_t$. Because of the very local nature of the model, a thermodynamic approach for solving the minimization problem has been used [27], in which spin changes are applied in a Monte-Carlo way until the system converges to its minimum. In each round, two cells are selected randomly on the interface between the fluid and the air. A random number is generated, and, if it is less than the probability of spin exchange, given by the Boltzmann law, the spins of two cells will be exchanged [24].

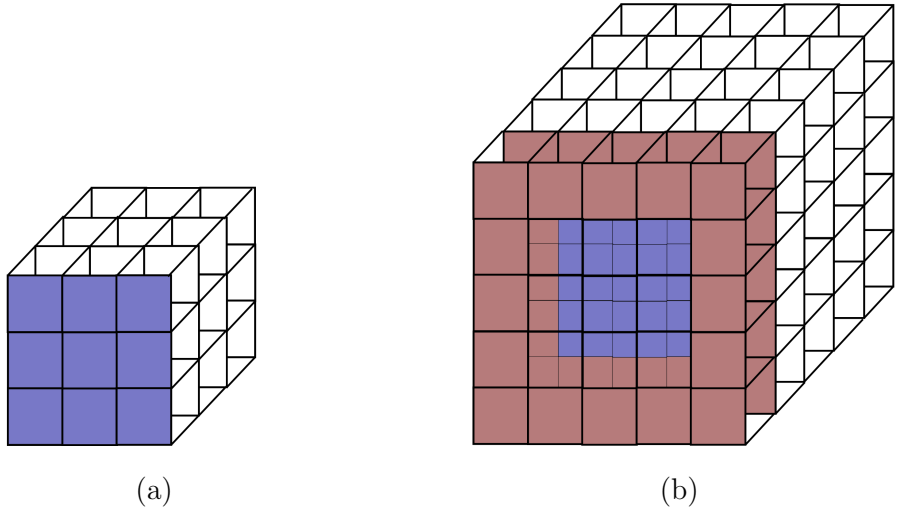

(a) (b)

Fig. 3. Schematics of neighbors around a typical cell in a 3D structure. a) First-layer neighbors. One side of the cells is colored. b) The second-layer neighbors referred to in section 6. In order to visualize the first-layer neighbors inside the structure, some of the sides are transparent.

The fluid propagation of this model is very local. Therefore, any marching or local evolution-based methods for finding the solution are slow, and local oscillations may dominate the evolution. This is the reason why thermodynamics-based approaches, such as spin exchange, have been used in the literature [24, 27].

In this work, we follow the same modeling methodology to describe the paper structure at the mesoscopic scale. In order to be able to simulate the transient and finite-volume flows, we modify the energy term to enable us to control the fluid volume within the computational space, and we develop a new numerical method to solve the associated optimization problem. The latter method is based on a modified heuristic optimization process based on a genetic algorithm (GA) [28] and a newly introduced finite-volume boundary condition. In this way, the numerical solution can be obtained for the finite-volume case at a reasonable computational cost, despite the very local nature of the modified Ising models. Although our model could be calibrated to simulate transient flows, in this work we focus only on the steady-state solution. This is because the final ink distribution is more important than the transient flow in the seepage of ink through paper. Although the transient seepage of ink over long periods of time (hundreds of years) is also of great interest in modeling the degradation of historical documents, an approximation of the transient flow is enough, and can be produced using our model. Also, in the case of an infinite volume of fluid, we will use the graph cuts approach [7]. This global optimizer is very fast and outperforms all local/marching-based methods.

It is worth noting that the porous medium has been modeled in another work as a set of containers of variable-size which are connected by pipes. The throughputs of the pipes are also variable, depending on the local parameters of the medium [23]. The model can be considered as a very local diffusion model which has been discretized at the size of pores. It is available as a part of the commercially available Pore-Cor Research Suite [23].

## 4  Problem statement

A paper document is available and a spatial distribution of the ink on the surface of that paper is given. The volume of ink is finite. The 3D distribution of the ink within the paper structure is required at mesoscopic scale. The paper is considered as a fiber structure. The construction of the fiber structure will be discussed in the next section, and the corresponding modified Ising model for representing this finite volume fluid problem will be given in section 6. The presence of ink in this discretized space is denoted by the field $\sigma$.

## 5  Fiber structure

To obtain the $\phi$ field that determines where we have solid material and where we have empty space, we first need to have a fiber structure. In this work, the fiber structure is created by placing fibers one by one on an imaginary surface on random coordinates and random directions. Each individual fiber is constructed of a chain of rectangular blocks. The blocks can rotate in parallel to the surface, up to a maximum degree, which determines the stiffness of the fibers. As fibers pile up on the surface, some of them bend, also this maximum degree. The width and the height of each block, and total fiber length, are other parameters of a fiber. A sample fiber structure is shown in Figure 4. As seen in the figure, the paper is positioned face-down in order to better visualize the internal volume of the fiber structure. This means that the gravity force is upward in this case. The ink that would be placed on the paper surface will be seen below the fiber structure in our model, and will propagate upward because of cohesive and adhesive forces, as well as the gravity force. For the sake of computation, the fiber structure and the air space between the fibers are discretized into a collection of equivolume cells. The cells are referred to using a single index, such as $i$, and the total computational space is denoted by $V$.

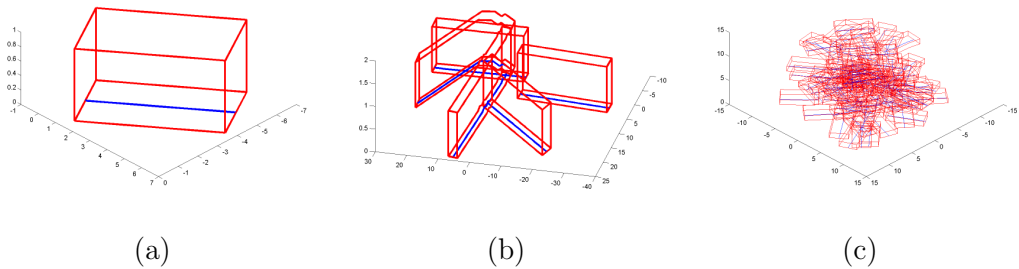

(a)          (b)          (c)

Fig. 4. An example of a fiber structure. The red lines show the edges of the blocks that compose the fibers, and each blue line is the guide line of a fiber, and passes through the center of the bottom surface of the blocks of that. For better visualization, the fibers are extended in the vertical direction. a) A single fiber. b) A collection of a few fibers. c) A fiber structure consisting of 100 fibers.

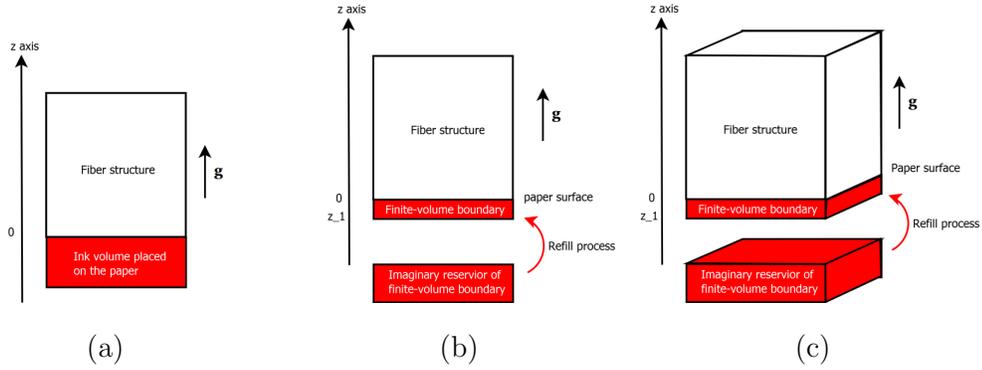

Fig. 5. A schematic diagram of the computational space used to simulate the ink seepage. a) A 2D view of the computational space before the finite volume boundary condition technique is applied. b) The same view after application of the finite volume boundary condition technique. Now, the computational space is independent of the ink volume. A refilling process ensures flow of the ink from an imaginary reservoir to the finite-volume boundary. c) A 3D view of the computational space.

# 6 Formulation

We follow the modified Ising model paradigm, presented in subsection 3, to develop a model which is able to describe finite-volume fluid flow at mesoscopic scale. First, we need to define the energy of the fluid-solid system. Three major terms contribute to this energy: 1) the gravity term, 2) the term accounting for the cohesion between the fluid molecules, and 3) the term accounting for the adhesion between the fluid and the solid molecules. Although the energy function (1) accurately describes the fluid-solid interaction between the ink and the paper structure, its mesoscopic nature makes it very local. Therefore, we need to add additional constraints to complete the description of the problem and take into account the other restrictions. We represent these constraints as additional energy terms. The total energy in the modified Ising model can be described as follows:

$$E_t = E_{t,0} + E_V = \sum_i \left( E_{g,i} + E_{c,i} + E_{a,i} \right) + E_V \qquad (2)$$

where $E_{t,0}$ and $E_V$ are the interaction energy and volume energy respectively. The volume energy stands for the finite volume constraint and is discussed later. Because of the high computational performance of our model, and to obtain higher accuracy, we also use the second-layer neighbors [24]:

$$E_{g,i} = G_g \hat{\sigma}_i z_i$$
$$E_{c,i} = -(c_1 \hat{\sigma}_i S_{1,i} + c_2 \hat{\sigma}_i S_{2,i})$$
$$E_{a,i} = -(A_0 \hat{\sigma}_i \phi_i + A_1 \hat{\sigma}_i F_{1,i} + A_2 \hat{\sigma}_i F_{2,i})$$

where $E_g$ is the fraction of energy corresponding to gravitational force; $G_g$ is the gravity constant; $\hat{\sigma}_i$ is the ink state of cell $i$, and $z_i$ is the vertical position of that cell. $E_c$ is the cohesive part of the energy between neighboring cells, where $c_1$ and $c_2$ are the coefficients of the first- and second-layer neighbors, and $S_{1,i}$ and $S_{2,i}$ are the ink states of the first- and second-layer neighbors:

$$S_{1,i} = \sum_{j \in N_{1,i}} \hat{\sigma}_j, \quad S_{2,i} = \sum_{j \in N_{2,i}} \hat{\sigma}_j$$

where $N_{1,i}$ and $N_{2,i}$ are the sets of cells that are the immediate neighbors and second-layer neighbors of cell $i$ respectively:

$$N_{1,i} = \{j | i \text{ and } j \text{ share a wall.}\}$$
$$N_{2,i} = \{j | i \text{ and } j \text{ share a } N_1 \text{ neighbor.}\}$$

Schematic diagrams of $N_{1,i}$ and $N_{2,i}$ are shown in Figure 3. Also, we use the notation $N_1$ for the set of all pairs $(i,j)$ such that $j \in N_{1,i}$ or $i \in N_{1,j}$.

$E_a$ is the adhesive part of the energy between the fluid and the solid, where $A_1$ and $A_2$ are adhesive coefficients for the first- and second-layer neighbors. The state of the solid material at these layers is provided by $F_{1,i}$ and $F_{2,i}$:

$$F_{1,i} = \sum_{j \in N_{1,i}} \phi_j, \quad F_{2,i} = \sum_{j \in N_{2,i}} \phi_j$$

where the state of the solid material in each cell is represented by $\phi_i$, the value of which is 0 or 1, depending on whether the cell is a free space or a part of a fiber. The self-adhesive coefficient is denoted by $A_0$. The second energy term (2), the volume energy part, is discussed in the following subsection. As can be seen from the definition of the energy, the energy is nonlinear in terms of the field $\sigma$. We discuss this point in more detail in section 7.

### 6.1 Volume energy: $E_V$

Total energy consists of both the interaction energy and the flow constraints. The most important constraint is the finite (and fixed) volume of ink (fluid). In other words, we need to force the incompressibility of the fluid via its fixed dimensionless volume:

$$V_{fluid} = V_{fluid,0} = \text{constant} \qquad (3)$$

where $V_{fluid,0}$ is the initial volume of ink and $V_{fluid}$ is the actual dimensionless volume, which can be expressed in terms of the $\hat{\sigma}$ field using $V_{fluid} = 1/2 \sum_{i \in V} (\hat{\sigma}_i + 1)$. This constraint can be added to the total energy in the form of another energy term, $E_V$:

$$E_t = E_{t,0} + E_V$$

$$E_t = E_{t,0} + \lambda \left( \sum_{i \in V} (\hat{\sigma}_i + 1) - 2V_{fluid,0} \right)^2 / (4V_0)$$

where $\lambda$ is a positive Lagrangian coefficient, $4V_0$ is used for the sake of normalization, and $V_0$ is the total volume of the computational space. The second term stands as an energy penalty term to enforce the volume constrain.

Before looking for a numerical method to solve the problem, we first reformulate the energy terms in such a way that the interaction terms, and also the range of the interaction, become more explicit.

We start with $E_{t,0}$:

$$\begin{aligned}
E_{t,0} &= \sum_{i \in V} (E_{g,i} + E_{c,i} + E_{a,i}) \\
&= \sum_i G_g \hat{\sigma}_i z_i \\
&\quad - \sum_i (c_1 \hat{\sigma}_i S_{1,i} + c_2 \hat{\sigma}_i S_{2,i}) \\
&\quad - \sum_i (A_0 \hat{\sigma}_i \phi_i + A_1 \hat{\sigma}_i F_{1,i} + A_2 \hat{\sigma}_i F_{2,i}) \\
&= \sum_i (G_g \hat{\sigma}_i z_i - A_0 \hat{\sigma}_i \phi_i - A_1 \hat{\sigma}_i F_{1,i} - A_2 \hat{\sigma}_i F_{2,i}) \\
&\quad + \sum_i (-c_1 \hat{\sigma}_i S_{1,i} - c_2 \hat{\sigma}_i S_{2,i}) \\
&= \sum_i (G_g z_i - A_0 \phi_i - A_1 F_{1,i} - A_2 F_{2,i}) \hat{\sigma}_i \\
&\quad + \sum_i \left( -c_1 \hat{\sigma}_i \sum_{j \in N_{1,i}} \hat{\sigma}_j \right) + \sum_i \left( -c_2 \hat{\sigma}_i \sum_{j \in N_{2,i}} \hat{\sigma}_j \right)
\end{aligned}$$

If we define $\sigma_i = 1/2(\hat{\sigma}_i + 1)$, i.e. $\hat{\sigma}_i = 2\sigma_i - 1$, we have:

$$E_{t,0} = \sum_{i \in V} D_{1,i} \sigma_i + \sum_{(i,j) \in N_1} (-4c_1) \sigma_i \sigma_j + \sum_{(i,j) \in N_2} (-4c_2) \sigma_i \sigma_j$$

where

$$\hat{D}_{1,i} = 2(G_g z_i - A_0 \phi_i - A_1 F_{1,i} - A_2 F_{2,i} + 2(26c_1 + 98c_2))$$

$\sigma$ is zero for empty cells, and 1 for cells containing fluid. The first term in $E_{t,0}$ is linear with respect to $\sigma$.

Now, the second term in the total energy can be reformulated:

$$V_0 E_V / \lambda = \left( \sum_{i \in V} (\hat{\sigma}_i + 1) - 2V_{fluid,0} \right)^2 / 4$$

$$= \left( \sum_{i \in V} \sigma_i - V_{fluid,0} \right)^2$$

$$= \left( \sum_{i \in V} \sigma_i \right)^2 - 2V_{fluid,0} \sum_{i \in V} \sigma_i + V_{fluid,0}^2$$

$$= \sum_{i \in V} \sum_{j \in V} \sigma_i \sigma_j - 2V_{fluid,0} \sum_{i \in V} \sigma_i$$

$$= \sum_{i \in V} \sigma_i^2 + \sum_{i \neq j} \sigma_i \sigma_j - 2V_{fluid,0} \sum_{i \in V} \sigma_i$$

But, $\sigma_i^2 = \sigma_i$

$$= \sum_{i \in V} \sigma_i + \sum_{i \neq j} \sigma_i \sigma_j - 2V_{fluid,0} \sum_{i \in V} \sigma_i$$

Therefore,

$$V_0 E_V / \lambda = (1 - 2V_{fluid,0}) \sum_{i \in V} \sigma_i + \sum_{i \neq j} \sigma_i \sigma_j$$

The total energy will be

$$E_t = E_{t,0} + E_V$$
$$= \sum_{i \in V} \hat{D}_{1,i} \sigma_i + \sum_{(i,j) \in N_1} (-4c_1) \sigma_i \sigma_j + \sum_{(i,j) \in N_2} (-4c_2) \sigma_i \sigma_j$$
$$+ \lambda/V_0 \sum_{i \neq j} \sigma_i \sigma_j + \lambda(1 - 2V_{fluid,0})/V_0 \sum_{i \in V} \hat{\sigma}_i$$
$$E_t = \sum_i D_{1,i} \sigma_i + \sum_{(i,j) \in N_1} V_{1,(i,j)} \sigma_i \sigma_j + \sum_{(i,j) \in N_2} V_{2,(i,j)} \sigma_i \sigma_j + \sum_{i \neq j} V_{3,(i,j)} \sigma_i \sigma_j$$
(4)

where

$$\begin{aligned}
D_{1,i} &= \hat{D}_{1,i} + \lambda(1 - 2V_{fluid,0})/V_0 \\
&= 2(G_g z_i - A_0\phi_i - A_1 F_{1,i} - A_2 F_{2,i} \\
&\quad + \lambda(1 - 2V_{fluid,0})/(2V_0) + 2(26c_1 + 98c_2)) \\
V_{1,(i,j)} &= -4c_1 \\
V_{2,(i,j)} &= -4c_2 \\
V_{3,(i,j)} &= \lambda/V_0
\end{aligned}$$

## 7 Solution of the variational problem and the experimental results

Before discussing the numerical method, we discuss the parameters. $c_1$ is selected as the reference parameter: $c_1 = 1$. To limit the influence of the second- and third-order terms, we assume that $98c_2 << 26c_1 \Rightarrow c_2 = c_1/8$ and $G_g z_{max} << c_1 26 \Rightarrow G_g = c_1/z_{max}$. Please note that 26 and 98 are the numbers of first- layer and second-layer neighbors respectively. In the adhesive terms, $A_0$ is the key parameter. We choose $A_0 = c_1/2$, in order to have stronger cohesive that adhesive behavior. The other parameters are $A_1 = 2/3 A_0$ and $A_2 = 1/2 A_1$ [24].

The optimization problem (4) is a nonlinear combinatorial optimization problem, which should be solved in a very large variable space. Figure 5 visualizes the computational space of the problem. The initial state of the ink is also shown in the figure. The solution is very far from the initial solution. As has been discussed in the related work section, local optimizers are usually used to move toward the global solution. However, because of the discrete nature of the problem, local forces (which are available at the microscopic scale) are not available to derive the system. For example, in [24], thermodynamics-based approach is used. Although these approaches are very helpful in evolving the simulation, the computational cost can be very high. Global optimizers, such as graph cuts [7], could be helpful. However, because of the ink-volume restriction in our problem, the basic requirement of this optimizer, which is that the range of interactions be limited to neighbors, is not met. Therefore, we only use this optimizer for the case of an infinite-volume fluid.

### 7.1 Solution via Global Optimization

As discussed above, the final ink distribution would be far from its initial distribution. Therefore, a global solver could result in a fast and economic solution. However, because of some restrictions in the problem, mainly the finite-volume restriction on the ink fluid and the mesoscopic scale of the energy function, the available global optimizers cannot be applied. The only exception is when the finite-volume restriction is removed. Below, a fast solution is presented using one of the powerful global optimizers, graph cuts [7].

In [7, 8],[1] the global solution to any optimization problem has been found if the problem can be expressed in the form of two terms, the first one being linear with respect to the variables, and the second perhaps being nonlinear (second-degree term), just between the neighbors:

$$E = \sum_{i \in V} D_i(\sigma_i) + \sum_{(i,j) \in N_1} V_{(i,j)}(\sigma_i, \sigma_j) \qquad (5)$$

where $D_i$ is a data penalty function, and $V_{(i,j)}$ is an interaction potential.

Comparing the general form of energy suitable for graph cuts methods (equation (5)) and our energy functional (4), we can see that the main terms that prevent us from applying the graph cuts method to this problem are the last two in (4) which count on $N_2$ neighbors and on almost all the cells in the computational domain respectively. The restriction posed by the first term could be removed by adding $N_2$ neighbors to the graph cuts formalism, which is what we did in our implementation here. However, the second term in (4) cannot be handled by this global optimizer, unless we set $\lambda = 0$ to drop this term, which means that there is no restriction on the fluid volume. In other words, we can solve the problem using the graph cuts method in the case of infinite fluid volume.

---

[1] http://www.mathworks.com/matlabcentral/fileexchange/21310, http://www.cs.ucl.ac.uk/staff/V.Kolmogorov/software.html

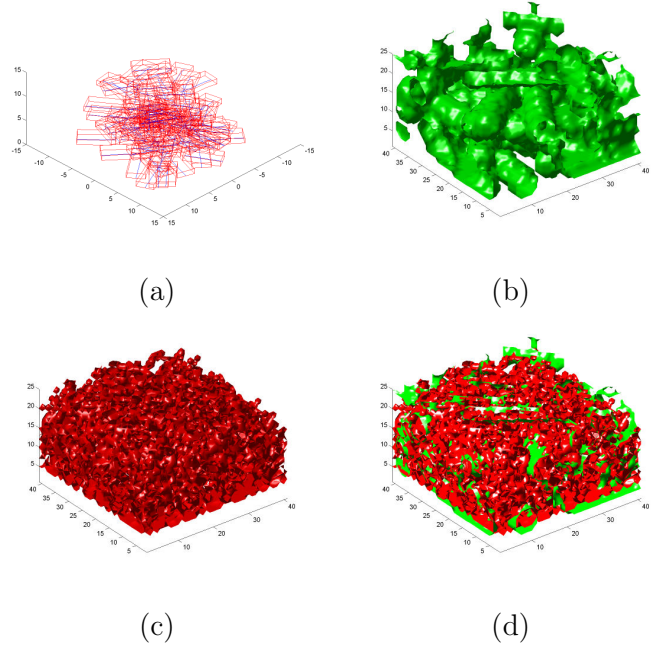

(a) (b)

(c) (d)

Fig. 6. A typical example of the ink propagation in the fiber structure under infinite-volume condition: a) The fiber structure. b) The $\phi$ field, which represents the fiber material. c) The $\sigma$ field, which represents the ink distribution. d) The ink and fiber together.

Figure 6 shows a typical example of the case of infinite fluid volume case. The fiber structure and ink distribution are shown. Because of the infinite-volume nature of this example, the ink fills the structure completely.

*7.2 Solution via a genetic algorithm*

A new method is developed in this subsection to solve (4) under the finite-volume condition ($\lambda \neq 0$). For this purpose, a modified genetic algorithm (GA) is introduced. The details of the method are given in Algorithm 1. It works, along with the finite volume boundary condition introduced in section 6, to reduce the computational cost. As can be seen from the algorithm, the method is quasi-linear, because it updates the nonlinear terms, $S_{1,i}$ and $S_{2,i}$, only on certain time intervals (100 iterations, in our work here). In this way, the energy functional will be linear in each GA iteration. Also, we limit the scope of the GA domain to the regions around the ink-air interface. The $\sigma_i$ variables in these regions are considered as genes, and then are composed as a chromosome of variable-size in each interaction. The binary nature of each $\sigma_i$ makes them intrinsically compatible with the GA approach. The cost function, which is linear, is low computational cost. A sample convergence curve of the energy is shown in Figure 7. The updating intervals of the nonlinear terms are shown with red lines. As can be seen in the figure, the convergence is smooth and stable.

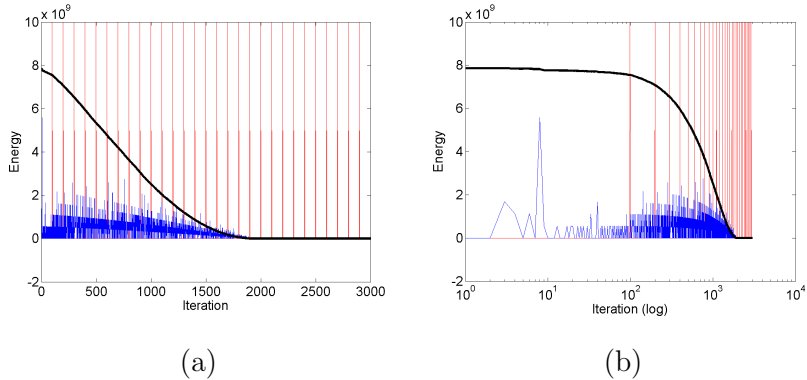

(a)  (b)

Fig. 7. The profile of the energy of the system, along with the computational time as a measure of convergence. a) The axis of the time is linear. b) The axis of the time is logarithmic. The red lines indicate the moments when the variables $S_1$ and $S_2$ are updated. The blue curve shows the spontaneous change in the energy along the computational time. As can be seen in the figure, the convergence is smooth and stable.

**Algorithm 1:** The modified genetic algorithm used to minimize the energy functional

1. Set $\sigma_i = 0$ for all cells with $z \geq 0$, $\sigma_i = 1$ for all cells with $z_1 < z < 0$ where $z_1$ is the depth of finite-volume boundary. If the volume of the finite-volume is bigger than the ink volume, $V_{fluid,0}$, fill just a part of finite-volume boundary of the same volume of $V_{fluid,0}$;
2. Initialize $S_{1,i}$ and $S_{2,i}$;
3. **repeat**
4.     **for** $k = 1$ *to* $100$ **do**
5.         Extract the cells that are on the interface of ink and air;
6.         Compose the genetic chromosome by aligning $\sigma_i$ variables of the extracted interface cells;
7.         minimize $E_t$ with respect to the created chromosome using genetic algorithm;
8.         Apply the optimal values back to $\sigma_i$ of the interface cells, and update the interface front;
9.         If the ink volume in the computational space is less than $V_{fluid,0}$, refill the finite-volume boundary cells ($z_1 < z < 0$);
10.     **end**
11.     Update $S_{1,i}$ and $S_{2,i}$;
12. **until** *The change in the energy is less than* $0.1$ *percent*;

In the first example, shown in Figure 8, a small amount of ink flows into a fiber structure. The finite volume of the ink is implemented by limiting the size of the finite-volume reservoir (see Figure 5(b)). The ink propagates into the fiber structure and settles around the fibers. We use the error in the ink volume ($\|V_{fluid} - V_{fluid,0}\|$) at the end of computations as a control parameter. This error is zero for the result of the method, and confirms the ability of our model to simulate finite-volume ink/paper interaction.

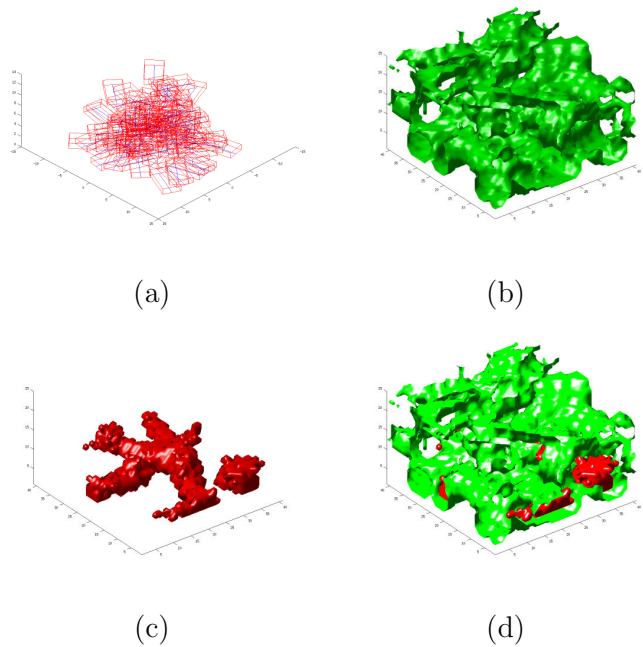

(a)          (b)

(c)          (d)

Fig. 8. The ink propagation in the fiber structure when the ink volume is finite: a) The fiber structure. b) The $\phi$ field, which represents the fiber material. c) The $\sigma$ field, which represents the ink distribution. d) The ink and fiber together.

In the second example, Figure 9, we study the extreme case of infinite-volume ink. Similar to Figure 6 of the graph cuts method, the ink fills the computational volume.

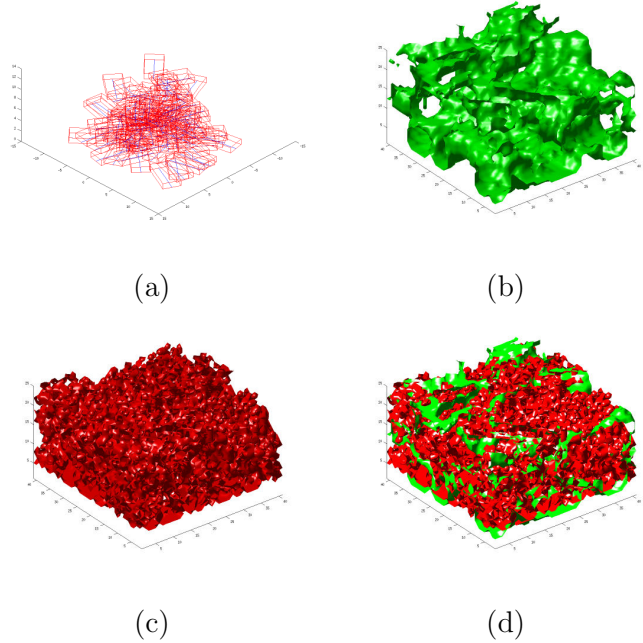

Fig. 9. The ink propagation in the fiber structure with a high volume of ink: a) The fiber structure. b) The $\phi$ field, which represents the fiber material. c) The $\sigma$ field, which represents the ink distribution. d) The ink and fiber together.

In another extreme case (Figure 10), we consider a very thin paper structure. As can be seen in the figure, the ink is mostly confined to the fiber area, despite the large amount of ink. When we increase the ink volume drastically, again the ink fills the whole computational space (Figure 11). However, it should be noted that this is a very rare case. Also, it is worth noting that the error in the ink volume is zero in all the cases. This is mostly because of a high $\lambda$ value (100 in this work), which enforces the finite-volume restriction.

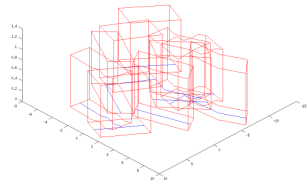 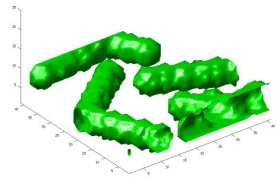

(a) (b)

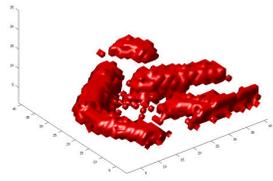 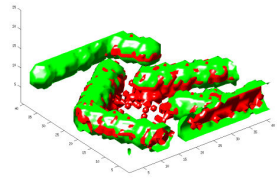

(c) (d)

Fig. 10. The ink propagation in the fiber structure when the fiber structure is very thin: a) The fiber structure. b) The $\phi$ field, which represents the fiber material. c) The $\sigma$ field, which represents the ink distribution. d) The ink and fiber together.

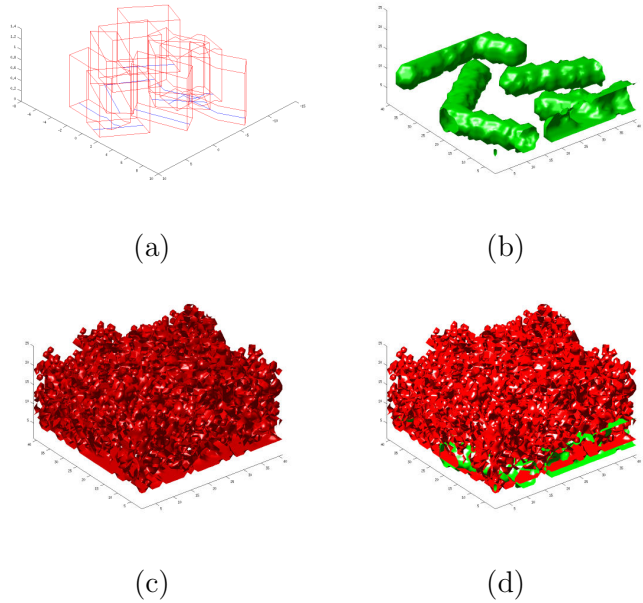

Fig. 11. The ink propagation in the fiber structure with a high volume of ink in a thin fiber structure: a) The fiber structure. b) The $\phi$ field, which represents the fiber material. c) The $\sigma$ field, which represents the ink distribution. d) The ink and fiber together.

## 8 Conclusion and future prospects

A new Ising model for the simulation of fluid seepage into a porous medium under the finite-volume fluid condition is developed. The model can be used to study fluid/solid interaction at mesoscopic scale. The computational cost is reduced by introducing quasi-linear behavior and a new finite volume boundary condition. Also, a modified genetic algorithm is introduced to solve the associated time-dependent optimization problem. The model is used to produce ink distribution in the fiber structure under a finite-volume restriction. We used the difference in the ink volume before and after simulation as the error measure, in order to assess the accuracy of the model. The results satisfied the volume restriction, showing zero error values.

In future work, we will use this model to obtain the diffusion coefficients of the macroscopic-scale models. This will help to avoid the need to perform large-scale simulations at mesoscopic scale, while at the same time transferring the physical properties of the mesoscopic interactions to the macroscopic scale.

## A  Macroscopic-scale models

Here, we review a few related macroscopic-scale models. These models work on continuum media, and therefore follow a different notation compared to the main text.

### A.1  2D Diffusion Model

Analysis at the macroscopic scale using generic diffusion-based models has been carried out in [18] by introducing a simple model consisting of two layers representing the paper material. The model ignores the fiber structure of the paper, which controls the very local flow of the ink. In that work, a diffusion-based model was developed to replicate the bleed-through effect in double-sided documents. It assumes that there are two independent distributions of ink on the recto and verso sides of a document: $s_{recto}$ and $s_{verso}$. The generic governing equation of that model is

$$\frac{\partial u}{\partial t} = \mathrm{DIFF}\left(u, s_{\mathrm{recto}}, c_{\mathrm{recto}}\right) + \mathrm{DIFF}\left(u, s_{\mathrm{verso}}, c_{\mathrm{verso}}\right) \qquad (\mathrm{A.1})$$

where $\mathrm{DIFF}(u, s, c)$ represents a diffusion process between the source layer $s$ and the target layer $u$ with the diffusion coefficients $c$. The target field $u$ is continuous and represents the 2D distribution of ink on the recto side of the document [18]. The diffusion coefficients define the physics converging the phenomena. For example, for the verso diffusion, we have:

$$c_{\mathrm{verso}} = \frac{d}{1 + (s_{\mathrm{verso}} - u)^2 / \sigma_b^2} \frac{1}{1 + s_{\mathrm{verso}}^2 / \sigma_{\mathrm{ink}}^2}, \qquad (\mathrm{A.2})$$

where the parameter $d$ is the ratio of the verso diffusion magnitude compared to that of the normal diffusion on the recto side. The parameter $\sigma_b$ controls the degree of ink seepage through the paper, and $\sigma_{\mathrm{ink}}$ is a general parameter that restricts diffusion for just the ink (which corresponds to near-zero values on $s$ and $u$). The coefficient $c_{\mathrm{verso}}$ has been designed in that work in such a way that it preserves the nonlinear nature of the ink seepage and also takes into account the spatial diffusion and dependency of the ink distribution within the paper material (for more details, see [18]). An example of the output of the model is presented in Figure A.1. The model is able to create bleed-through signatures in a realistic way. It is worth noting that the model is not directly inferred from the physical ink seepage phenomenon, but instead, the diffusion coefficient is selected as a closed form based on the empirical understanding of the phenomenon. However, the formula for the diffusion coefficients is the main concern, and it should be verified by the simulation models at lower-level scales.

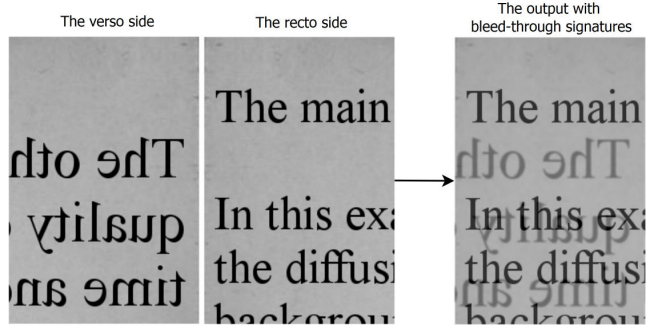

Fig. A.1. An example of bleed-through effect generated using the diffusion-based model [18]. Left side: The recto and verso side ink distributions before the bleed-though effect. Right side: The final output of the diffusion model. It simulates the bleed-through interference.

A.2  *3D macroscopic models*

A better description of ink seepage could be obtained using 3D models. Under slow flow conditions, the flow in the porous substrate can be described by Darcy's equation (derived from the Navier-Stokes equations in cylindrical coordinate system) [3, 12]:

$$\frac{1}{r}\left(r\frac{\partial p}{\partial r}\right) + \frac{\partial^2 p}{\partial z^2} = 0 \qquad (A.3)$$

where $p$ is the fluid pressure, and the velocity components are given by

$$u_r = \frac{-\phi}{\mu}\frac{\partial p}{\partial r}, \quad u_z = \frac{-\phi}{\mu}\frac{\partial p}{\partial z}$$

Here, $\mu$ is the dynamic viscosity, and $\phi$ is the permeability. These equations are appropriate when the porous skeleton is saturated. However, in the case of printing onto a dry porous substrate, computation of the flow into the skeleton is also necessary. This may be achieved using an approach which computes the position of the metal front in a casting simulation [10]. Numerically, this is described by an additional equation, the hyperbolic transport equation:

$$\frac{d\hat{\sigma}}{dt} + u_r\frac{\partial \hat{\sigma}}{\partial r} + u_z\frac{\partial \hat{\sigma}}{\partial z} = 0 \qquad (A.4)$$

where $\hat{\sigma}$ is a scalar quantity to be transported. For example, $\hat{\sigma} = -1$ may represent a dry position and $\hat{\sigma} = 1$ may represent a wet position. Usually, a numerical solution of the model is obtained using the finite element method [29]. The model is macroscopic, and does not consider individual fibers in the paper structure. The material properties are described using averaged variables, such as $\mu$. The model is able to follow the transit flow of ink/water through the substrate.

## A.3 Flow-based models

The non-zero flow in a porous medium could be modeled as a flow of incompressible viscose fluid, as follows [1]:

$$\frac{\partial}{\partial_t}\mathbf{u} + (\mathbf{u} \cdot \nabla)\mathbf{u} = \frac{-1}{\rho}\nabla p + \nu\nabla^2\mathbf{u} + \mathbf{g} \tag{A.5}$$

where $\mathbf{u}$ is the velocity vector, $\mathbf{g}$ is the gravity constant vector, $\nu = \mu/\rho$ and $\mu$ is the dynamic viscosity. In the case of stationary flow with no friction, we have the Stokes equation:

$$\nabla p - \rho G_g = \mu\nabla^2\mathbf{u} \tag{A.6}$$

where $G_g$ is the gravity constant, and the space changes of velocity are considered negligible ($(u\nabla)u = 0$). In the literature, this model has been solved using Lattice-Boltzmann hydrodynamics [1, 33, 30]. By definition, the model assumes a stationary flow. This means that transitional flow cannot be covered by this model, and also the amount of fluid should be infinite. Moreover, this assumption forces a periodic boundary condition. Therefore, the model is not suitable for the study of ink propagation through the paper. However, there are some models based on Navier-Stokes equations adapted to free-boundary fluid flows.